\documentclass[11pt,letterpaper]{article}
\usepackage[margin=1in]{geometry}
\usepackage{amsmath,amssymb}
\usepackage[expansion=false]{microtype}
\usepackage[hidelinks]{hyperref}
\usepackage{setspace}
\title{The Threshold Theorem in Watts\\ \large\itshape Fault Tolerance as a Question About Objective Probability}
\author{Amit Hagar\thanks{HPS Department, Indiana University, Bloomington, IN 47405.
Email: \texttt{hagara@iu.edu}. This paper extends the argument the author made together with Giuseppe Sergioli in \emph{Counting steps} (arXiv:1101.3521, 2011). AI assistance was used for accelerating research, source assembly and verification, and grammatical editing; the text and all claims are the author's, written and then checked by hand.}}

\begin{document}
\maketitle

\begin{abstract}
\noindent In 2011 Hagar \& Sergioli proposed a new interpretation of objective probability in deterministic physics. On it, the probability of a physical state supervenes on the resources, energy over time, required to realize it from a given state, relative to the resources available (arXiv:1101.3521). The motivation for that paper was an objective alternative to QBism, the view that sees quantum probabilities as subjective degrees of belief. Here I apply this interpretation to a more practical subject matter: fault-tolerant quantum computing (FTQC). Under the resource-bounded measure Hagar \& Sergioli proposed, the threshold theorem becomes a claim about classification: it asserts that error-corrected logical states belong to the class of relatively cheaply realizable states, whose probability remains near 1 as the machine grows. The theorem originally derived this claim from a resource inventory that was partial, and left out five resources consumed by error correction: calibration of a drifting device, decoding within the correction cycle, coherence as a finite time budget, entropy flush through fresh ancillas, and the control power that suppresses coherent noise below threshold in the first place. These entered the original derivation at zero price. Here I translate the feasibility of FTQC into a measurable quantity, watts per decade of suppressed logical error (a decade, in the engineer's usage, being one factor of ten in the error rate). I then show that the published record already contains its first data points for this translation, and I state the two measurements that would settle the question empirically. The three-decade debate on FTQC, conducted so far as an exchange about noise-model assumptions, turns out to be, under this interpretation, a quantitative dispute about a single object: the resource-bounded objective probability of the target logical states.
\end{abstract}

\section{Introduction}

The threshold theorem is one of the most foundational documents of the quantum computing industry, second only, perhaps, to Shor's algorithm. It states that {\em if} the error rate per physical component lies below a critical constant, arbitrarily long quantum computations can be performed with arbitrarily small logical error, at a resource overhead growing only polylogarithmically in the target accuracy \cite{aharonov1997,knill1998,kitaev1997}. Every valuation of a quantum computing startup and every announced qubit count or runtime in the quantum computing industry's roadmaps or projections rests on this theorem.

The debate over its applicability to physical devices is as old as the theorem. Even before the theorem existed, and immediately after Shor announced his algorithm, Landauer \cite{landauer1995} and Unruh \cite{unruh1995} raised concerns about dissipation and decoherence. That skepticism, in a way, pushed the mathematicians to come up with error-correction that eventually led to the theorem. A few years later, Alicki and collaborators argued that its noise assumptions conflict with the Hamiltonian derivation of open-system dynamics \cite{alicki2006}. Kalai has famously pressed a statistical version of the same doubt \cite{kalai2020}. The skeptic list is short, but each argument carries weight and led to modifications of the noise conditions. The optimists answer, correctly, that the theorem is a theorem: given its assumptions, the conclusion follows. The pessimists answer, also correctly, that the assumptions describe no device anyone has built.\footnote{I set aside the recent ``below threshold'' results \cite{willow2024,bluvstein2024,he2025}: they are evidence of breakeven at small distance, and reading them as evidence of full-scale fault tolerance conflates a measured quantity, $\Lambda$ at distances 3 through 7, with an extrapolated one, $\Lambda$ held constant to distance 30. Currently those extrapolations rest on the noise model that the theorem favors, which is exactly what is at stake. Worse, the extrapolated object is not even a stable measurement: Vezvaee et al.\ show that the reported $\Lambda$ comes from a one-parameter fit assuming cycle-independent, unital, SPAM-free noise, that their own data violate all three assumptions, and that the resulting number moves by 10--15\% with the choice of fit \cite{vezvaee2026}.} Three decades in, the exchange on the feasibility of FTQC has the structure of a border dispute in which each side maps the territory in its own grid of coordinates.

Since I entered the field in 1993 I have held that the feasibility question is an empirical matter of fact, one settled only when the machine is built. This view stands in contrast to the view that ties the existence of FTQC to the universality of quantum mechanics or to the failure of the Extended Church-Turing Thesis \cite{Aaronson2004}. My view is less metaphysical. I simply observe that quantum mechanics may be universal alright, but thermodynamics might exact its toll eventually, and rather than philosophical, the practical question is whether we can squeeze some useful quantum computation before that toll. Until the issue is settled, this paper proposes a common coordinate system so that the dispute could at least be maintained on rational grounds regardless of one's optimistic or pessimistic personality. 

In 2011 I introduced, together with Giuseppe Sergioli, an interpretation of objective probability for deterministic physics in which the probability of a physical state is a measure of the resources required to realize it, relative to the resources available \cite{hagar2011}. Section 2 summarizes the construction without returning to the motivation that fueled it. Section 3 shows that under this measure the threshold theorems of the 1990s amount to a classification claim, an assertion about which of two complexity classes the targeted logical state occupies, argued from a resource inventory that the intervening years of experiments have shown to be partial. Section 4 itemizes those unpriced resources, with the published record as evidence that each is now measurable, and some of it already obtained. Section 5 draws a corollary for the choice between active correction and passive protection, a choice made by the industry at the beginning of the century. Under the new interpretation, that choice can now be evaluated with hindsight. Section 6 locates this proposal among the existing literature. Section 7 details thirteen objections the argument anticipates, each with its reply. I welcome attempts to rebut those replies or to marshal more objections. Section 8 states the two measurements that would decide the question, one of which can be performed this year on existing hardware with a power meter.

Two remarks on the scope of this paper before we begin. First, I make no claim that fault tolerance fails, and I have no stake in whether it succeeds. I claim only that (1) whether it succeeds is an empirical question about a resource curve, that (2) the curve has a precise definition under an interpretation of objective probability published in 2011, and that (3) the curve is currently unmeasured. Second, the probability interpretation itself is independent of its application: it stands as a physical reading of the probabilities that appear in the formalisms of statistical and quantum mechanics whether or not the reader accepts my accounting of error correction. The independence runs the other way as well: the two measurements of Section 8 are stated in watts, error rates, and hours of downtime, so a reader who rejects the probability interpretation wholesale can still run those measurements. In short, the meter I propose here and the measure we proposed fifteen years ago need not consult each other.

\section{Probability as resource-bounded realizability}

The interpretation of objective probability Sergioli and I proposed in 2011 rests on five working hypotheses: the dynamics of physical systems is deterministic; there is a difference in kind between dynamical evolutions whose computational cost grows polynomially with system size and those whose cost grows exponentially \cite{hartmanis1965}; the resources of the universe are bounded, so the set of possible evolutions is finite; the time--energy uncertainty relation limits the resolution of energy differences, so this set is also discrete \cite{aharonov2002}; and interactions are local. I believe these assumptions are innocuous for those who treat physical processes governed by Schr{\"o}dinger's equation as computations, which, for the target readership of this paper, is an easy ask.

On these assumptions, the notion of objective probability is that of a transition: every physical process can be analyzed as a computation from one state to another, a sequence of discrete steps each consuming energy and time. Fix a system of dimension $n$, an energy budget $E$, and a time allowance $t$. The available power $\mathrm{Pw} = E/t$ bounds the number of computational steps $\#(\mathrm{Pw})$ that can be completed between any two states, and this bound in turn selects, from the finite set $S$ of dynamical evolutions that could realize a target state from the initial one, the subset $A$ whose cost fits the budget. The probability of realizing the state is the ratio
\begin{equation}
P = \frac{|A|}{|S|},
\end{equation}
a quantity that satisfies Kolmogorov's axioms \cite{hagar2011}.\footnote{For one family of rotations on a single qubit the resulting measure reproduces the Born rule as a state overlap \cite{hagar2011}.} $P = 1$ means the system is at the target and the transition costs nothing. $P = 0$ means no algorithmic process reaches the target at any finite budget. Between the extremes, $P$ is a distance measure. It quantifies how far the system is from the target in the only currency nature accepts. While probability is dimensionless, a ratio, every probability statement in the theory, which is a statement about the physical world and not about our degrees of belief, carries a price, and the price is denominated in watts.

One clarification forestalls a possible charge of circularity. The steps counted are steps of the actual physical dynamics of the system at hand, approximated to arbitrary accuracy; the measure is therefore machine-relative. A quantum evolution may consume fewer resources than any classical evolution realizing the same state, and the 2011 paper reads the violations of Bell's inequalities as marking exactly this difference in cost \cite{hagar2011}. Pitowsky \cite{Itamar} is for me the canonical reference for the idea that the difference between classical and quantum mechanics is a difference in a probability structure (Boolean versus non-Boolean) and not in metaphysics. That difference is the basis for the entire edifice of quantum computing, and I do not reject it. On the contrary, I embrace it. A framework that counted only classical steps would rule against quantum computing by definition; mine prices each machine on its own dynamics, and this is what permits the comparison proposed in Section 8.

Two consequences of the construction matter here. First, in 2011 we constructed a warm-up model and applied the same accounting to the space of all possible states of a large system. The assumption of bounded resources led to the assignment of a collective measure near 0 to the exponentially costly states and a collective measure near 1 to the polynomially cheap states, and the assumption of locality led to near-uniform weights within the latter. But while this partition of states into \textit{Poly} and \textit{Exp} with measures near 1 and near 0 was assumption-based, which class a given state of interest belongs to remains a contingent, empirical question. We noted this contingency in 2011 and I emphasize it here again. Second, in quantum mechanics the time--energy relation puts a bound on the accounting: predicting a system requires knowing its Hamiltonian. But knowing the Hamiltonian requires estimation, and estimating an unknown Hamiltonian to precision $\Delta H$ costs time $\Delta t$ with $\Delta t \Delta H \geq 1$ \cite{aharonov2002}. So the bound can be lowered by better estimation but it cannot be removed.

\section{The threshold theorem as a classification claim}

I will now translate the threshold theorem into this currency of transition probability as a distance measure with a price tag in watts. A logical qubit held at error rate $\varepsilon$ is a state of probability $1 - \varepsilon$, and under the 2011 measure that probability supervenes on the resources spent holding it. The theorem asserts that below a critical physical error rate, the resource cost of maintaining the logical state at error $\varepsilon$ grows as $\mathrm{polylog}(1/\varepsilon)$ and polynomially in the computation size. In our terms, the theorem asserts that the fault-tolerant logical state belongs to \textit{Poly}. Above threshold, correction is a net negative. It consumes more than it repairs, so the required-resource curve turns exponential, and the state falls into \textit{Exp\/}. As stated above, \textit{Exp\/} has probability near 0. The threshold is thus the crossing point between the two classes of the 2011 partition, hence the classification claim.

Stated this way, the original theorem which was a statement in circuit combinatorics now becomes a conjecture about the shape of a cost curve, with the noise model as its free parameter. Back then, the original derivation priced some resources explicitly: physical qubits ($2d^2 - 1$ per logical qubit at distance $d$ for the surface code \cite{fowler2012}), gate counts, syndrome measurements. And it took other resources at zero: the theorem's device does not drift, its decoder is instantaneous, its coherence window is unbounded, its supply of fresh cold ancillas is free, and its noise is already weak enough to correct without any control power spent beating it down first. Whether the logical state remains in \textit{Poly} once every resource is counted and is on the books is exactly the question the contingency clause of Section 2 reserves for experiment.

I therefore propose the native unit of the debate on the feasibility of FTQC: {\bf watts per decade of suppressed logical error}, one decade reducing $\varepsilon$ from, say, $10^{-3}$ to $10^{-4}$. Parts of the industry, for different reasons, have already begun measuring quantum advantage in energy terms \cite{ETH}. My interpretation from 2011 recasts computational complexity in physical resources and as such may appeal to those supporting the recent currency drift. Some readers may prefer joules per logical operation, or energy per shot. They can substitute freely since the unit family is interchangeable, and the claim defended here is only about the shape of the curve, which is invariant under the choice. The theorem predicts this curve is nearly flat. The five inventory items of the next section are five independent arguments that the physically realized curve steepens, and, decisively, five demonstrations that the curve is now measurable, with some data points already obtained.

\section{The unpriced inventory}

The five items below are the ones for which a published record already exists, but the list is open.\footnote{Other entries that come to mind are the cost of magic state distillation, or the training cost and the inference cost per cycle of AI control agents. These too scale between generations.}  Note, however, that whether or not the list becomes exhaustive is immaterial to the meter I propose in Section 8. That experiment will tally the total watts regardless of what one calculates in advance. Only the theorem requires completeness.

\subsection{Calibration}

The devices the theorem speaks of have a fixed, known noise channel. But every physical device has a Hamiltonian that drifts, and the bound of Section 2 applies to the apparatus even where syndrome extraction cleverly avoids applying it to the data: syndrome circuits learn which error occurred without learning the encoded state, but the circuits themselves must be calibrated. Calibration is Hamiltonian estimation, and estimation is priced at $\Delta t \Delta H \geq 1$, repeatedly, for the lifetime of the machine, because the parameters do not stay estimated.

One response is to calibrate continuously: estimate the noise model directly from syndrome statistics and update it in a Bayesian fashion as the device drifts, with the decoder retuned to the current model. A genuine engineering answer that relocates the bill rather than cancelling it. Downtime becomes continuous classical inference, running every cycle, inside the same latency budget that already binds the decoder, and it is read by the same meter.

The published record already sets a budget for this line item. Google's frequency-optimization paper states that the calibration problem is non-convex, highly constrained, time-dynamic, and expands exponentially with processor size, spanning roughly $3N$ interdependent frequencies (where $N$ is the number of qubits). It also concedes that re-optimizing all gates of a processor when outliers are detected is unscalable from a runtime perspective \cite{klimov2024}. Another paper, the Willow ``below threshold'' result, reports in its drift test three drift-recalibration events, one between every four of sixteen experimental runs across a 15-hour campaign, targeting qubit frequency and readout drift, after a preparatory step that forecasts two-level-system defect frequencies to survive the initial calibration \cite{willow2024}. A subsequent Google paper states the situation quite plainly: the current solution is to terminate the entire quantum computation for recalibration, which is incompatible with the long runtimes of future quantum algorithms \cite{rlqec2025}. Third-party measurements quantify the cost of not paying for this line item: on Rigetti hardware, a single gate left uncalibrated for eight hours raises the distance-3 logical error rate by 41.6\% (single-qubit) and 135.5\% (two-qubit) \cite{caliqec2025}. An architecture subfield has formed since 2024 whose sole purpose is paying this cost in situ \cite{caliscalpel2024,caliqec2025,reloqate2026}, which is the strongest available evidence that the line item is real.

I must acknowledge the counter-reading. In the same Willow drift test, the error-suppression factor held at $\Lambda = 2.18 \pm 0.07$ across the 15 hours, and runs immediately after recalibration were not appreciably better than runs just before. This is the strongest published evidence for the optimist: a 72-qubit code is robust to the drift levels present, over 15 hours. But it is evidence about a 72-qubits processor and 15 hours. The question the resource measure asks, how the recalibration cost scales when the configuration space expands exponentially and the algorithm runtime is 8 hours \cite{gidney2021}, the same 15-hour window now containing three drift-recalibration events, is left exactly open. Both readings should be in the record, and only the cost curve decides between them.

\subsection{Decoding}

Correction is conditioned on a classical computation that must finish inside the correction cycle, because a decoder that falls behind accumulates a syndrome backlog, and the computation time blows up exponentially in the algorithm's non-Clifford depth \cite{terhal2015}. Willow's real-time decoder reports an average latency of $63 \pm 17\,\mu$s at distance 5, against a cycle time of $1.1\,\mu$s, sustained over a million cycles, and the authors note that the latency scales with the code size, underscoring the need for further optimization \cite{willow2024}. The decoder, in our terms, is a second computation running on a second budget, whose cost enters the denominator of the logical state's probability and whose scaling with distance is, by the authors' own sentence, an open engineering front. The original threshold derivations assumed decoding completes within the correction cycle at no cost to the logical error. The latency race I just described is what that assumption actually costs in the lab.

The decoder's budget is now also a published number. The cooling budget of the 4-kelvin stage of a dilution refrigerator is roughly 1.5\,W \cite{krinner2019}, and transmitting raw syndromes from that stage to room temperature costs about 0.92$\mu$W per syndrome bit, so the full budget supports only about one thousand distance-7 logical qubits before a single correction has been computed \cite{cryozip2026}. The engineering response to this budget is a hardware literature running since 2021: cryogenic decoders (2.78 $\mu$W per decoding unit, against a budget of about 1\,W at 4\,K in that paper's units \cite{qecool2021}), predecoders and syndrome compressors \cite{cryozip2026}. The result is a second subfield, alongside the calibration one, whose sole purpose is fitting into a power budget. Yet another item the original threshold theorem never mentioned.

\subsection{The coherence battery}

As Unruh first observed in 1995 \cite{unruh1995}, every operation of the corrected machine, including the correction, must complete inside the coherence window of the device. That window is a hard physical time budget: the Willow processors reported a mean $T_1$ of $68\,\mu$s, a distance-7 logical lifetime of $291\,\mu$s. They also reported correlated error bursts of unexplained origin roughly once per hour. These further impose a current logical error floor of $10^{-10}$ in repetition codes \cite{willow2024}. The benchmark algorithm the industry cites for utility runs eight hours \cite{gidney2021}. A correction (defined in physical time) is one that must arrive before the window closes. That deadline is a resource, and consuming it has a price the abstract theorem never entertained. On a more abstract level, a control-theoretic formulation of this very point has recently been developed independently \cite{pighin2026}. It defines minimal steering times under bounded drive amplitudes, with the decoherence time as the horizon of every control problem.

\subsection{Entropy flush}

Error correction is entropy flush: syndrome extraction moves the entropy generated by noise into ancillas, and the ancillas must arrive fresh and cold, at the cycle rate, forever. Each erasure is priced by Landauer's bound at the fundamental floor and by cryogenic engineering at the practical one. The holding cost of a logical qubit is therefore a power, energy per unit time, which is precisely the quantity $\mathrm{Pw}$ that indexes the 2011 measure Sergioli and I proposed. Full-stack estimates that take this accounting seriously find that the power bill of a fault-tolerant machine at algorithmic scale can change by three orders of magnitude with the design parameters, gigawatts under poor choices against megawatts at the optimum. This optimum may be reached only through joint optimization of the entire stack \cite{fellous2023}. The same group's earlier analysis, which only priced the photons at the qubits and left attenuation and cryogenics off the books, had yielded nanowatts \cite{fellous2021}. The closest thing to a measurement in the literature is a RAND working paper that prices a cryptanalytic computation end to end: extrapolating from existing dilution refrigerators to roughly 6.25 W per physical qubit (about the power consumption of one LED light bulb), a machine breaking one RSA-2048 key would draw about 125 MW and consume about 890 MWh. In the authors' own words: ``nearly 10 times the power consumed by the Summit supercomputer, about the power consumption of a Boeing 747 aircraft in flight, and about a quarter of the power produced by a typical coal-fired power plant.'' The authors then estimates that at the 2022 average U.S. industrial price, breaking one public key would cost \$64{,}000 of electricity \cite{rand2023}. The same study reports that for the other factor in its estimate, the average power per qubit, it could not find any prior quantitative estimates at all. No hardware paper in the field currently reports the quantity all of these estimates are circling around, which I propose as the decisive quantitative measure: the measured power drawn per decade of suppressed logical error.

\subsection{Error suppression}

The items above are consumed by the correction cycle. This last resource is consumed before the surface code runs at all, and it is the one that most directly converts watts into suppression. On any device whose noise is coherent or non-Markovian, the surface code cannot reach its threshold unaided; the residual noise must first be beaten down by dynamical decoupling (DD), a train of control pulses applied to every idle qubit. Vezvaee et al.\ report that their heavy-hex surface-code memory requires up to roughly 484 single-qubit DD pulses per cycle to fill the idle gaps, and that these pulses do most of the work: without DD the logical error rate for the $\Lambda$-favored basis is far higher, and the apparent scaling advantage of the larger code can vanish once DD is applied to the smaller code as well \cite{vezvaee2026}. Those pulses are microwave power drawn at the control racks, the same instrument column the meter reads. So here the accounting is exact rather than inferred: the suppression the field quotes is bought with control power, and the amount of control power scales with the number of idle gaps, which grows with circuit depth, which grows with code distance. This is the cleanest single instance of the paper's thesis: a resource that the reported metric credits on the suppression side while leaving its cost off the meter, made visible the moment one counts the pulses. It also carries a warning that belongs in the measurement of Section 8, and I return to it there: a suppression figure is meaningful only when both codes being compared are DD-optimized, because otherwise the improvement is partly an artifact of an under-suppressed baseline \cite{vezvaee2026}.

\section{A corollary: active and passive protection}

To demonstrate the utility of the probability measure we proposed in 2011 in a broader context, consider the following example. In 1999 the field, not an industry yet, was faced with two strategies for noise resilience, the active approach based on concatenated codes, and the passive approach based on the notion of noiseless subsystems, or decoherence-free subspaces. The probability measure sets two different price tags on the two approaches, and the difference explains a choice the field made without stating it. Active correction, the threshold program, pays an operating cost: power, ancillas, decoding, and recalibration, every cycle, scaling with machine size, for the lifetime of the computation. Passive protection, the program of Zanardi, Rasetti, Lidar, Whaley and others \cite{zanardi1997,lidar1998}, pays a capital cost: a full characterization of the noise structure and an engineered symmetry, purchased once at design time. The active program could be entered on a single cheap number, the physical error rate. In contrast, the passive program could not be entered without an expensive portrait of the noise. The field entered where entry was cheap. The resource-bounded measure integrates the whole contract rather than just the entry price, and under it a low entry fee attached to a divergent operating cost can be seen as the worse purchase. The 2024 demonstration of dynamically generated decoherence-free subspaces on superconducting hardware \cite{quiroz2024}, arriving twenty-five years after the theory, suggests the comparison may have closed prematurely and can be reopened in the common currency I propose here, although I doubt the industry will be the one to reopen it. The worry that passive error avoidance confines quantum computing to bespoke tasks, those whose structure admits a protective symmetry, evaporates once one notices that the active approach, for all its advertising of general-purpose computing, has so far produced four or five algorithm families. The menu was always short, and a small set of verifiable wins beats an unbounded set of unverifiable ones.

\section{Relation to prior work}

The empirical part of this paper exists in the literature and is peer-reviewed. The resource-constraint results of Fellous-Asiani, Chai, Whitney, Auff\`eves and Ng \cite{fellous2021,fellous2023} are one example. They show that scale-dependent noise under bounded resources yields a maximum attainable accuracy, which further grounds the energetics of full-stack machines. Those papers compute the cost curve that this paper points at, but, naturally, as engineers, they leave the notion of probability untouched. My aim here is to supply the missing half: an interpretation under which the cost curve is not merely an engineering criterion but also, maybe more important to the debate on the feasibility of FTQC, the physical content of the probability statements the threshold theorem makes. This interpretation, proposed by Sergioli and me, predates the energetics literature by nine years (arXiv:1101.3521, January 2011, against arXiv:2007.01966, July 2020).

The famous critiques of the theorem's assumptions \cite{landauer1995,unruh1995,alicki2006} are ancestors of Section 4 but propose no probability measure. Kalai's program \cite{kalai2020,kalai2026}, whose fully depolarizing noise conjecture (FDNC) specifies the structure that correlated noise would take, is the closest the skeptics have gotten to a full-fledged structural attack. Kalai's instrument and mine are complementary in a precise sense: he conjectures the structure while I set a price tag on its suppression. His own discussion concedes that the conjectured channel is never observed directly but must be inferred through tomography complicated by calibration, drift, and gauge freedoms, and that the noise it predicts ``may well be produced, partly or largely, by mechanisms already known to experimentalists and regarded as engineering difficulties'' that ``may not be merely engineering obstacles.'' Those difficulties are exactly the line items of Section 4, which means the meter offers his conjecture an indirect test that requires no tomography at all. If Kalai's FDNC is true, the machinery that launders correlated noise into effective independence works harder every generation, and the marginal slope climbs. On the other hand, a flat, fully priced slope would be evidence against his conjecture.

The resource theories of quantum information \cite{veitch2014,howard2017,danageozian2022} count rigorously but abstractly, in monotones without watts or devices, and they, too, leave probability untouched. Pighin's control interpretation of quantum advantage \cite{pighin2026} performs the same accounting as the 2011 measure in the opposite direction, defining advantage rather than probability by resource-bounded realizability on named machines; his construction and mine are complementary halves of one program.

The surrounding energetics literature has meanwhile grown into a program of its own. Here are some examples: the call for a quantum energy initiative \cite{auffeves2022}, the measured energetics of a single qubit gate \cite{stevens2022}, the efficiency analyses of full machines and data centers \cite{ikonen2017,jaschke2023,martin2022}, and a framework in which quantum computation holds an energy-consumption advantage \cite{meier2025}. Another example is a thermodynamic analysis of one (so far, unpriced) entry in this list, the Landauer heat of syndrome erasure, finds a phase transition between a bounded-error phase where cooling mitigates it, and an unbounded phase of runaway heating. It also places current parameters in the bounded phase, conditional on present hardware capabilities being maintained at scale \cite{bilokur2024}. That condition is in effect what Section 8 proposes to measure (among all the other unpriced conditions that keep the noise model i.i.d.). And a fridge that mitigates this heat is not the same as a marginal slope that stays flat, because while a bounded phase constrains the temperature, the marginal watts per decade remain untouched by it. 

While this literature sets a price on the full stack or on parts of it, none of it ties that price to the probability of the logical state or to the assumed noise model, which is the linkage I propose here. For the experiment in Section 8, the record from this literature is promising but also incomplete. Data points such as power figures, decoder latencies, and error rates per distance already exist in print, but no published dataset yet joins them on one machine across code distances.

\section{Objections and replies}

I have made my claims and I now turn to the objections they might attract, each with its reply. I also welcome rebuttals of my rebuttals, and additional objections if they appear.

\begin{enumerate}

\item \textit{Discreteness smuggles epistemology.} The time--energy relation limits distinguishability, but it makes no claim on existence, so the discreteness hypothesis of Section 2 allegedly imports epistemology into an ontology the framework was motivated to purge. My reply is, and was always, operationalist. Take any two Hamiltonians. If you can't distinguish between them with a finite-resource process, then that means they do not differ. Or consider my claim that infinite precision is the $P = 0$ case of the measure, or the claim that discreteness is the measure applied to the space of dynamics, which is the operational corollary of the no-difference point above. All are operationally justified with no recourse to degrees of belief. Even the relativity to a budget leaves the measure objective, and the budget is a fact about system and apparatus. So no need for degrees of belief either. The stricter objection, that no rigorous time--energy uncertainty relation exists because time is not an operator, is answered by Aharonov, Massar and Popescu \cite{aharonov2002}. They derive the estimation cost without a time operator and this estimation cost is the only use I make of the relation. This also answers the Bertrand-style worry about graining: operationalism need not entail conventionalism. The partition of the space of dynamics is set by the budget's resolution, and refining it below distinguishability adds cells that cannot be told apart by any physical process.

\item \textit{A Kolmogorov measure cannot yield quantum statistics.} Bell, Kochen--Specker, and, closer to home, Pitowsky's correlation polytopes \cite{Itamar} show that no single classical probability space reproduces the quantum correlations, so if the 2011 measure aspires to the Born rule in general, it must fail by the very results I call canonical. My reply is that the measure we proposed is machine-relative by construction: the set $S$ is fixed by the dynamics, the resource budget, and the experimental arrangement, so the sample space is contextual. Within one arrangement the structure is Boolean and the measure is classical, but when one moves across arrangements the events do not embed in a common Boolean algebra, which is precisely the structure Pitowsky identified as quantum probability. So the measure actually instantiates his theorems.

\item \textit{The physical Church--Turing thesis (not the polynomial, extended one, mind you) is assumed, but not proved.} So the hypercomputation literature will observe, correctly. Guilty as charged. Assumed indeed, as a contingent working hypothesis. And the reason is operationalism, again. The known counterexamples, which my mentor Pitowsky highlighted in his seminal work from the 1990s, e.g., non-globally-hyperbolic spacetimes and ill-posed initial data \cite{pourel1989,hogarth1994}, are realized in no lab.\footnote{I note that Pitowsky gave these examples not to argue for the physical possibility of hypercomputation but to warn against a naive translation of the notion of computation from mathematical logic to physics. We actually heed his warning with our 2011 measure because we define computation on the physics we have, machine by machine, budget by budget.} The objection is also self-undermining in the present context. The threshold theorem is a statement about Turing-computable circuits, so anyone defending fault tolerance has already signed the physical Church--Turing thesis. If quantum computing required physical processes outside the reach of any resource accounting, the optimist would have a hard time winning the debate.

\item \textit{The Born-rule result is numerology.} The reproduction of the Born rule reported in Section 2 covers one family of rotations on a single qubit, and was only given in 2011 as a plausibility argument. The construction is qualitative, and its convergence to the observed quantum probabilities in full generality remains the open conjecture of the 2011 paper. Nothing in this paper depends on it. My argument here uses only the resource accounting, and it survives regardless of the Born-rule remark.

\item \textit{The reading will come out flat.} On any single present-day machine the watts-per-decade reading will come out flat, because the power budget is dominated by fixed costs. A dilution refrigerator draws its tens of kilowatts whether the code runs at distance 3 or distance 7, and a flat reading on one fridge would vindicate nothing. This is correct, but note that I priced it in advance: the quantity defined in Section 8 is a slope, {\em the marginal power per decade as code distance and machine size grow across generations\/}. It is not a single reading, so the fixed cost drops out of such a marginal slope.

\item \textit{The falsifier is one-sided.} Branding me as a skeptic (I am not) guarantees the question of motive will be raised, so let me be explicit about what I propose. The measurement of Section 8 is one the optimist should want: a marginal power curve that stays flat as the machines scale would be the strongest evidence for FTQC ever published, stronger than any below-threshold demonstration, because it would price the whole enterprise rather than just the entry fee. And note that the falsifier is concrete on both sides, because a marginal slope that remains flat across two successive hardware generations spanning an order of magnitude in qubit count, with recalibration downtime and decoder compute on the meter, settles the inventory objection of Section 4 against me. I also read the error bars against myself: a slope statistically indistinguishable from flat, with everything on the meter, settles it for the theorem as well. On the other hand, a slope that climbs across the same two generations, with each generation's engineering improvements already on the books, settles it in the inventory's favor, and closes the escape by which a rising cost is attributed forever to immature engineering. In other words, I am actually proposing the experiment that can prove me wrong, faithful to my being a (non-naive) empiricist. Once both sides have signed on, the last section is a contract, and a cheap one at that.

\item \textit{Category error.} My pedigree as an outsider may trigger readers to judge the measurement proposal of Section 8 to belong in a hardware journal and the interpretation of Section 2 in a philosophy journal (yes, I am a philosopher, among other things). This combination, however, describes the paper's subject, and is not a defect. The empirical half of the paper is already in the physics literature by the community's own standard \cite{fellous2023,fellous2021,rand2023}, while the interpretive half has been on the arXiv since 2011 (and was published in an obscure Italian journal that no longer exists). So the question spans the two literatures because the resources do so too. And besides, if the feasibility of FTQC can be checked today with a power meter, why should one care about pedigree? 

\item \textit{This is not new; the energetics literature already sets a price on the stack.} Yes, it does. I credit it in Section 6. But note what that literature does: it computes the energy bill under an {\em assumed} i.i.d. noise model. The repairs are delegated to better engineering, which in turn requires cost estimations.\footnote{Google's own conclusion: ``the path to fault tolerance will be built not only on better hardware but on more intelligent control'' \cite{rlqec2025}. Intelligent control is still control: the RL agent that replaces the calibration halt is itself a computation, with a training cost and an inference cost per correction cycle, a new entry together with Snake. Its price at scale is once again the quantity I propose to measure in Section 8.} My proposal runs in the exact opposite direction. Under the 2011 measure the cost becomes {\em evidence\/}, because the burden of the maintenance materializes the noise assumption: if the noise is as the theorem assumes, the marginal slope stays flat. If correlations persist, the machinery that twirls them into effective independence works harder every generation and the slope climbs. So I agree that the energetics camp sets a price in watts on fault tolerance, but the meter proposed here actually allows us to cross-examine it. The linkage between wattage cost and the noise hypothesis is what no prior watt figure was asked to entertain, and it is what is at stake for the threshold theorem (and, if you are an operationalist, also for its funding).

\item \textit{The calibration bill is classical combinatorics, not physics.} The bill of Section 4 is a product of three factors: the number of parameters to estimate, which grows with the machine; the bound per estimation, set by $\Delta t \Delta H \geq 1$; and the repetition rate, which is set by the drift. Cleverness can negotiate only the first factor. It can't help with the second, which is just quantum mechanics doing what it does, universally. Now, I acknowledge there is a known escape from the second factor, and it is instructive to see why it is unavailable here: Atia and Aharonov showed that Hamiltonian fast-forwarding permits exponentially precise energy measurements, so a naive universal reading of the time--energy principle is a misconception \cite{atia2017}. But fast-forwarding is a quantum computation, and its precision guarantee presupposes that this computation runs with sufficient accuracy. In other words, it presupposes the fault tolerance whose price we are trying to audit. Zeroing the calibration bill with fast-forwarding is like doing a balance transfer between your own two credit cards. Sometime in the future, that $0\%$ APR rate becomes $29\%$. In a world where the marginal slope climbs, the operational content of the time--energy bound is broader than the ideal circuit model suggests.\footnote{The escape and its closure were added in v2, prompted by \cite{kalai2026}, where the same point about the operational content of the time--energy bound is made independently.}

\item \textit{Snake answers the calibration objection.} Google's frequency-optimization paper offers its answer to the calibration problem, the Snake optimizer, projected by simulation to remain effective at a distance-23 logical qubit \cite{klimov2024}. The projection is a simulation of the heuristic, and the projection is itself an entry in the unpriced column: what Snake will cost at that scale, in runtime and in power, is exactly the quantity Section 8 asks to be measured.\footnote{The optimizer was introduced in 2020, validated on the 53-qubit Sycamore supremacy processor, and projected to scale favorably with qubit number \cite{klimov2020}; four years later their own numbers frame this line item: a runtime budget of 0.5 hours chosen for compatibility with operating large surface codes, a measured $\sim3.6$ seconds of optimization runtime per qubit that reaches $\sim1.4$ hours at $N = 1057$. They also add a stitching rescue that restores the budget by splitting the processor into disjoint regions, scaling sub-linearly in principle, while its behavior at seams, where outliers risk amplification, rests at target scale on simulation \cite{klimov2024}. Their own words.}

\item \textit{$\Lambda$ already measures this.} Why a new quantity, when the suppression factor $\Lambda$ can simply be tracked at each new code distance? Because $\Lambda$ is the numerator of the 2011 ratio but with no denominator. It compares logical error rates between successive distances, but is measured {\em after} calibration, with decoder compute, recalibration downtime, and entropy flush {\em all off the meter\/,} so the two quantities can diverge. ``$\Lambda$ holding above 2 while the marginal power per decade climbs'' is a different statement than just ``$\Lambda$ holding above 2,'' and that divergence is the whole dispute about the feasibility of FTQC. And note that the proposal isn't trading one asymptotic for another, because watts per decade is a slope read off finite hardware, without taking any limit or extrapolating, while the extrapolation that currently all vendors are making from small-distance $\Lambda$ out to distance 25 or 30 assumes the noise stays independent across an exponentially growing spacetime volume. I have argued elsewhere in public that the correlated-burst floor of $10^{-10}$ reported in Section 4 already puts this assumption in question. And the standing reply, that the noise is measured all the time, survives only while the measuring apparatus stays in the instrument column: the calibration loop is what manufactures the effectively independent noise $\Lambda$ is measured against, so every hour of drift tracking invoked as reassurance is a line item on the bill which is currently unreported as if it were free. The dynamical-decoupling result of Section 4 sharpens the same point on the numerator side: Vezvaee et al.\ show that a reported $\Lambda$ can be manufactured by comparing a well-tuned large code against a baseline that was not itself DD-optimized, so that the improvement is an artifact of the comparison rather than a property of the hardware \cite{vezvaee2026}. The suppression that this manufactured $\Lambda$ credits is bought with hundreds of control pulses per cycle, which are watts on the meter. That $\Lambda$ can be inflated by leaving those watts uncharged is precisely the divergence this objection concerns, now demonstrated inside a single experiment.

\item \textit{Some codes are already robust to correlated noise.} Single-shot error correction achieves correctness under noise classes that break the standard assumption, and its robustness against temporally correlated noise has been established \cite{bombin2015,bombin2016}; whether it survives the sharper correlated models now being proposed \cite{modi2026} is not yet settled.\footnote{Both this entry and the discussion of continuous calibration in Section 4 were prompted by correspondence with Kavan Modi.} But note how the robustness is bought: these schemes demand substantially more physical qubits, more syndrome extraction, and heavier decoding. Error correction under correlated noise is thus purchased with resources, which is the claim of this paper. The watt meter does not care which column the purchase is being tallied in; it reads the total.

\item \textit{The partition is asymptotic, so no finite slope classifies the curve.} A polynomial with a large exponent can lie above an exponential with a small base at every size anyone will ever build, so the division into \textit{Poly} and \textit{Exp} is meaningful only in the limit, and a measured slope, mine included, classifies nothing. Yes, we acknowledged this ourselves already in 2011. But the 2011 measure is finite by construction, probability relative to a budget, so the question was never the curve's class ``in the limit'' but its price at the scale the machine inhabits. And this paper names that scale: the eight-hour benchmark of Section 4 \cite{gidney2021}. A marginal slope that climbs across the generations approaching that scale defeats the promise at every size that will ever be built, whatever the asymptotic class turns out to be, while a slope that stays flat there vindicates FTQC in the only sense that carries cash value (operationalism, yet again). In any case, the asymptotic escape belongs to the theorem, and the meter I propose never relied on it and does not have to. And note that my concession cuts symmetrically: if a rising finite slope cannot prove \textit{Exp}, a flat one at distance 7 cannot prove \textit{Poly}. That is exactly the point I made in my rebuttal of the previous objection, from the other side.
\end{enumerate}

\section{Two measurements}

Finally, I show how the three-decade dispute on the feasibility of FTQC reduces, under the probability measure proposed here, to two curves that are yet to be published.

First, {\bf watts per decade of suppressed logical error, on one machine, as the code distance grows}. The numerator is read from a power meter at the dilution refrigerator, the control racks, and the decoder workstation; the denominator is $\Lambda$, already reported to two decimal places \cite{willow2024}. One condition on that denominator, which the record does not yet enforce, follows from Section 4: both codes entering the ratio must be separately optimized over dynamical decoupling, since a $\Lambda$ read against an under-suppressed baseline credits an improvement the hardware did not make \cite{vezvaee2026}. This condition costs the proposal nothing, because the control power that the optimization consumes is already inside the numerator; charging it there and requiring it in the denominator simply closes the loop. Every quantity in this measurement exists today but the measurement is absent. The proposed quantity is a slope: the marginal power per decade as code distance and machine size grow across generations, with control lines, decoder compute, and recalibration downtime included in the meter. Downtime enters as energy, since facility power drawn during hours in which no logical result is produced is part of the price of the results that follow. And the within-generation slope is already accessible: Willow ran distances 3, 5, and 7 on one processor in one campaign \cite{willow2024}, so the marginal cost per distance step is measurable on a single device, with cross-generation data supplying the size axis. Two successive generations are enough to settle the direction of the marginal power, and while later generations will keep testing its shape, the verdict does not need to wait for them. The theorem predicts the slope stays nearly flat at scale, while the inventory of Section 4 predicts it climbs, with the calibration and entropy terms as the leading corrections. 

Second, {\bf the symmetric classical curve}. Advantage and feasibility claims currently compare a priced quantum machine against an unpriced abstraction, the best known classical algorithm. That abstractness led to no end of trouble in the debate on quantum advantage in the last eight years. So rather than leaving it abstract, I propose that the same resource accounting, drift, controls, bounded power, be applied to the classical competitor on the same task. That application yields a second curve, and the feasibility of FTQC becomes the empirical statement that the two curves cross, at a machine size and a power budget that the plot itself names.

Both measurements are cheap relative to any roadmap that the industry has produced. And neither requires new theory, because the ``theory'' has been on the arXiv since 2011. What the field has instead produced since then is a body of assumptions about noise, an exchange between optimists and pessimists that settled nothing, because assumptions are the one resource which is even cheaper than watts. So instead of continuing to argue about noise, I end with the observation that the noise model is the first assumption the meter proposed here converts into a hard currency: i.i.d.\/, under this accounting, stops being a premise and becomes a maintained burden that has a cost, and if correlations persist, their price appears on the bill whether or not anyone models them.

\end{document}